\documentclass[prb,preprint]{revtex4-2} 

\usepackage{amsmath}  
\usepackage{amsfonts} 
\usepackage{graphicx} 
\usepackage{amsbsy}
\usepackage[utf8]{inputenc}

\begin{document}


\title{Is the equivalence principle useful for understanding general relativity?}


\author{Peter C. Aichelburg}
\email{aichelp8@univie.ac.at} 
\affiliation{Faculty of Physics, Univ. of Vienna, Austria}



\date{\today}

\begin{abstract}
The Equivalence Principle (EP) is at the heart of General Relativity (GR), tested in many aspects (see C.M. Will \cite{Will}). It is often used to discuss qualitatively the influence of gravity on physical phenomena. But can this be made more precise? We compare clock rates, frequency shifts, light deflection and time delay in simple static spacetimes to the analogous phenomena seen by accelerated observers in Minkowski space. In contrast to previous  studies, we do not assume that  the gravitational field is weak and see, as we proceed, how the field is constrained by the EP. Special care is taken that results are only observer-, but not coordinate-dependent. By this we clarify some of the issues raised in the literature and show which gravitational effects can and which cannot be simulated by acceleration. The paper may also serve as a contribution for critical discussions on the implications of the EP. 


\end{abstract}
\maketitle
\section{Introduction} 
It is well known that Einstein's ``most fortunate thought of his life" \cite {Einstein2002} that a body in free fall is weightless, led him to the idea that gravity is a geometrical property of space and time. 
In his paper of 1911, ``On the influence of gravity on the propagation of light''
\cite {Einstein1911} Einstein postulates that a uniform gravitational field is equivalent with regard to \it all physical  processes  \rm to a system in uniform acceleration with respect to an inertial frame. Later \cite{Einstein1912}, he refers to this as  ``equivalence hypothesis'', and it has become known as equivalence principle (see comment\cite {Nordstroem}).
Despite the fact that in almost all textbooks on GR (e.g. see \cite{MTW}) one finds a discussion on the EP, 
its exact formulation, interpretation and usefulness for understanding physical phenomena in the presence of gravity, were and still are, under debate \cite{Hamilton,Desloge, Munoz,Ferraro, Moreau} including a very skeptical comment by Synge \cite{Synge}.
 There exist a number of different formulations of the principle and efforts to classify and relate them (see e.g. Di Casola et.al \cite{Casola} and references therein. For an analysis of the EP in quantum mechanics, see Giulini \cite{Giulini}).

The aim of the present paper is not to contribute to this debate but to stick to Einstein's original  version of 1911 and discuss its role for understanding GR not only on a qualitative level  but also some of its mathematical consequences.
 The EP is often used for a pedagogical introduction to GR. That relative rates of clocks are affected by gravity and that gravity influences the propagation and the frequency of light can be inferred qualitatively by making use of it.  Einstein himself, in the cited paper, applied the EP to derive the redshift formula and to obtain an expression for the bending of light passing near the sun on the basis of Newtonian gravity. Here we discuss whether these arguments can be made more precise by making the following assumptions:
\begin{itemize}
\item Gravitation is described by a  metric tensor theory. 
(We do not require Einstein's equations to hold in general, except when discussing local light bending where the Schwarzschild spacetime is considered as an example). 
\item Test particles, in our case clocks and detectors, move along time-like geodesics and light rays along null geodesics.
This assumes that all the internal degrees of freedom as well as the self-gravitating field can be neglected.

\item In absence of gravity Special  Relativity is valid.
\end{itemize}
We restrict our discussion to static gravitational fields and see to which extent the influence of gravity on clocks and light rays can be reproduced by going to an appropriate accelerated, i.e. non-inertial reference frame in the absence of gravity. It is generally agreed that the EP can hold strictly only locally, when tidal forces can be neglected (but see comments on this issue by Ohanian \cite{Ohanian}). Nevertheless, and in contrast to most of the articles on the EP,  we do not make use of a weak field approximation, nor do we require, except for light bending, confinement to small regions of spacetime. Rather, as we go along, we see how the gravitational field will be restricted in order to match the effects seen by an accelerated observer in the absence of gravity. Not surprisingly, at the end it turns out that if the EP is to be valid without restrictions, the Riemann tensor of spacetime has to vanish  and the gravitational field is spurious. 
Although most of the presented calculations are elementary, we make use of differential geometry and assume spacetime to have a Riemannian structure. The main drawback in most of the publications dealing with the EP is that arguments rely on specific coordinates. Here  we take care that results are only observer-, but not coordinate-dependent. This presentation clarifies some of the issues raised in previous papers and also gives a new perspective which might be useful when addressing the EP and its consequences.

The paper is organized as follows: For the discussed phenomena we first repeat the qualitative argument before turning to its mathematical formulation.
In Sec.II we start by comparing non-inertial clocks to an inertial clock at encounter. Specifically, in a static gravitational field (metric) 
we consider a freely falling clock passing two stationary clocks and compare their rates, Fig.1. By virtue of the EP, the stationary clocks are replaced by accelerated clocks in Minkowski space and their rates compared to an inertial clock. (A similar ``operational'' approach was followed by Anderson and Gautreau \cite{Anderson}).  We discuss if and how it is possible to reproduce the relative rates under the influence of gravity by suitable acceleration in Minkowski space.  
 (Arms and Serna \cite{Arms} proposed an experiment along this line). 
Since we want to mimic the effect of a static gravitational field, the corresponding acceleration should be uniform, that is constant in proper time in the rest frame of the observer, but not necessarily equal for all observers. Two cases are discussed: rigid motion, where spatial distances in the momentary rest frame  stay constant and, alternatively, clocks which undergo equal acceleration.
 Before turning to the frequency shift of light rays propagating in a static field in Sec.III,   
 we summarize results and make comments on the restrictions imposed on the gravitational field in order to match the effects seen by accelerated observers. Especially we argue that the ratio of static clock rates even in a uniform i.e. gravitational field of constant force,  cannot be reproduced in Minkowski space.  
In contrast to time dilation, the Doppler shift for light is of first order in the relative velocities, and it is interesting to see how this compares to the gravitational frequency shift. Again, we discuss this for both cases, rigid and equal acceleration.
Sec.IV is devoted to light bending. Here we consider the picture of Einstein's famous elevator, not freely falling, but at rest in a static  gravitational field and compare the results with a uniformly accelerated cabin without gravity. By local we mean that we do not integrate the orbit of the ray but rather look at the bending of the ray at a given point,
More precisely we calculate the curvature of the projection of the ray onto the subspace orthogonal to a static observer. We show that the spatial curvature of the ray is equal to the acceleration which a static observer feels at that point. Using the same acceleration for the cabin in Minkowski space, one cannot expect to reproduce the whole orbit, however one does obtain the same local curvature of the ray as seen by an accelerated observer. This holds true for the rigid as well as for the  equally accelerated cabin.
A number of papers have considered the role of the EP for derivation of the light deflection  \cite{Ferraro,Moreau}. One of the issues discussed is to disentangle the contribution of the static acceleration from that due to the curvature of 3-space.  Our calculation shows that there is a contribution from the space curvature even to the local bending. But at the same time the spatial curvature gives a contribution to the acceleration of static observers. However both contributions are such that the above statement remains true. Thus the local bending does depend on the curvature of 3-space, but so does the static acceleration.  In  this sense the local bending can be deduced from the EP and, as pointed out by Ehlers and Rindler  \cite{Ehlers} is a purely kinematic effect. 
Finally, in Sec.V, we touch upon the question whether the time delay experienced by a light ray in the presence of  a gravitational field, the Shapiro effect \cite{Shapiro}  has its   counterpart in Minkowski space. For the sake of simplicity we consider a radial ray bouncing back from a mirror located deeper in the gravitational potential and show that a similar (but not equal) effect can be seen by an accelerated observer in Minkowski space.
\section {RELATIVE CLOCK RATES}
In a general spacetime consider a geodesically moving clock C intersecting the trajectories (worldlines) of two non-inertial clocks A and B.
 Let ${\bf u}_A$, ${\bf u}_B$ and ${\bf u}_C$ be their corresponding 4-velocities.  The scalar products of the 4-velocities of A and B at the encounter with clock C are related to their corresponding relative velocities by the $\gamma$-factor (which in standard \it local \rm Minkowski coordinates is $\gamma (v) = (1 - v^2)^{-{1/2}}$).
\begin{equation}
\label{scalar1}
({{\bf u}_A}\cdot{\bf u}_C) = -\gamma_{A,C},   \qquad({\bf u}_B\cdot{\bf u}_C) = -\gamma_{B,C},
\end{equation}
where the scalar product is defined as $ ({\bf u}\cdot{\bf w}) =u^a  g_{ab} w^b $, and we use signature  $(-+++)$ for the metric and set the velocity of light $c = 1$.
(Here and in the following the scalar products of the 4-velocities are to be  evaluated at the crossing of the worldlines.)

An observer associated with clock  C with proper time $\tau_C$ sees clock A running at reduced rate (time dilation) according to
\begin{equation}
\label{rates1}
d\tau_A = \gamma_{A,C }d\tau_C = -({\bf u}_A\cdot{\bf u}_C)d\tau_C
\end{equation}
and similar for B with respect to C.
Taking the rate of C as reference, the ratio between the rates of A and B is given by

\begin{equation}
\label{ratio1}
 \frac{d\tau_A}{d\tau_B} = \frac{\gamma_{A,C}}{\gamma_{B,C}} =
\frac{({\bf u}_A\cdot{\bf u}_C)} {({\bf u}_B\cdot{\bf u}_C)}
 \end{equation}
 (Note that this is not to be understood as a differential but as the ratio of the rates as seen by an observer moving with clock C.) This operational procedure of assigning  relative clock rates for clocks at different points in spacetime may look ad hoc at first sight. However, for the special gravitational setting we have in mind, we show that this relation is independent of the specific geodesic of C.
More specifically we look at the situation of a freely falling clock C in a static spacetime passing two stationary clocks A and B  situated at different heights (see Fig. 2).
For simplicity we restrict spacetime to 1+1 dimensions and write the metric as  
\begin{equation}
 \label{metric1}
  ds^2= -f(r)dt^2 + h(r)dr^2,
\end{equation}
where $f(r)$ and $h(r)$ are arbitrary positive functions of the coordinate $r$, except that we require
$ f '(r) > 0$ , so that the freely falling clock moves along the negative $ r$-direction (note that any spherically symmetric static metric in 4 spacetime dimensions  can be written in this form when restricted to the radial direction, see e.g.(\cite{Wald}).
\begin{figure}
 \centering
\includegraphics[scale=0.5]{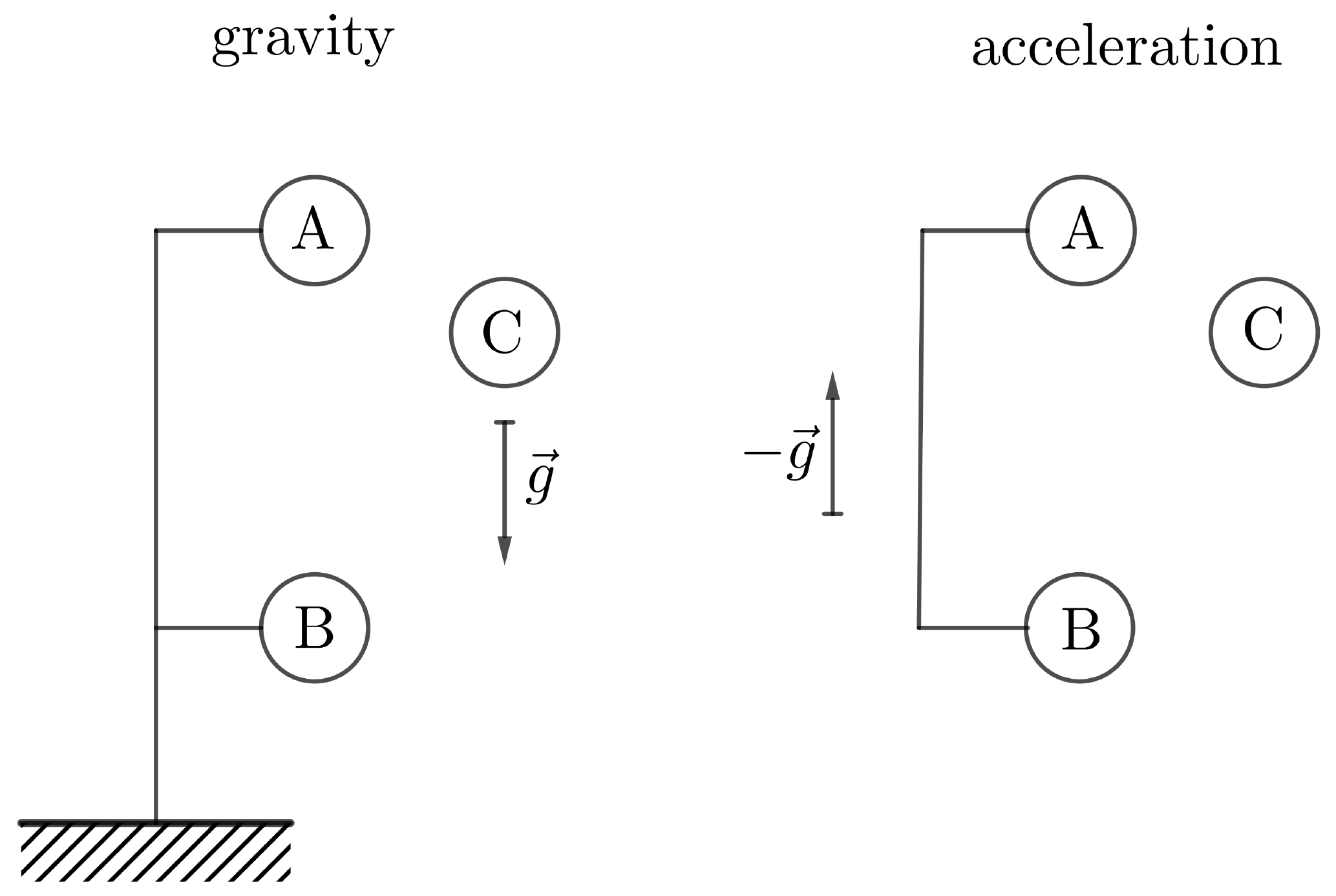}
 \caption{FALLING CLOCK,  the rates of two static clocks, A and B are compared to the rate of a freely falling clock C,
 (l.h.s.). Analogous, the accelerated clocks A and B in Minkowski space pass by an inertial clock C where rates are compared, (r.h.s.).
}\label{HereIsYourLabel}
\end{figure}
\begin{figure}
\centering
 \includegraphics[scale=0.6]{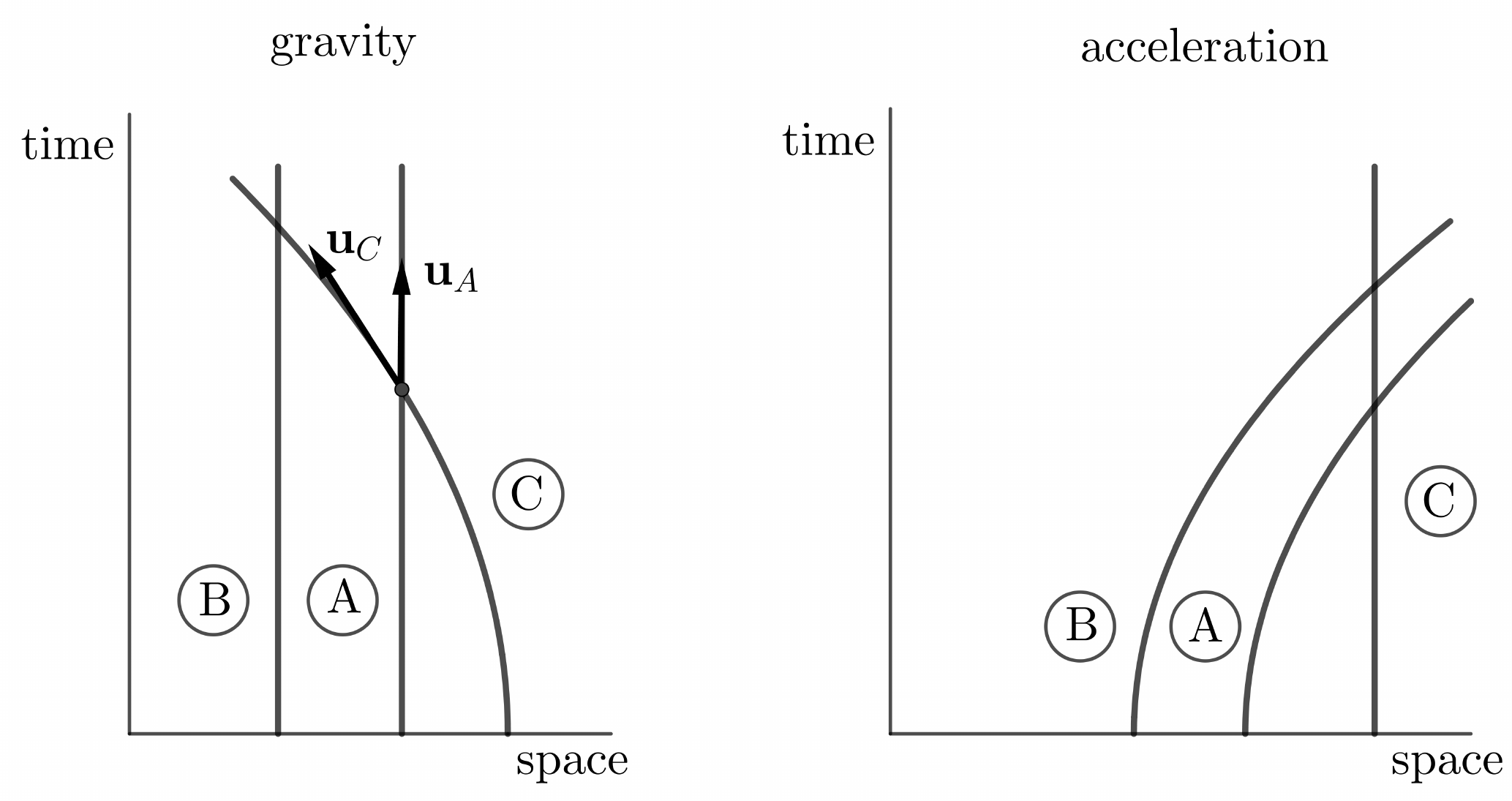}
 \caption{CLOCKS IN SPACETIME, static clocks A and B in gravity in comparison to accelerated clocks in Minkowski space. C is an inertial reference clock in both cases.
}\label{HereIsYourLabel}
\end{figure}
The clocks A and B are at rest, say at $ r_A$ and $r_B$  and we calculate the scalar products
 $({\bf u}_A \cdot {\bf u}_C)$ and $({\bf u}_B \cdot {\bf u}_C)$ at $r_A$ and $r_B$ respectively.
Using proper time along the trajectories $ t(\tau)$ and $r(\tau)$ implies 
 \begin{equation}
 \label{norm}
  ({\bf u}\cdot {\bf u)}  = - f(r)\dot{t}^2 + h(r)\dot{r}^2  = -1
\end{equation}
where the dot means differentiation with respect to proper time. The 4-velocities  (we use the term  ``4-velocity'' although here we restrict to  1+1 dimension)                                                                   of the static clocks are
\begin{equation}
  \label{velA,B}
  {\bf u}_{A,B}  =     
\begin{pmatrix}
  f^{-1/2}(r_{A,B})\\
 0\\
 \end{pmatrix}
\end{equation}  
The freely falling clock C moves along a geodesic. Time independence of the metric implies that the scalar product of $\bf u_C$ with the (Killing) vector field $\pmb{\xi}$, where ${\xi}^a ={\delta^a}_t $,  is constant along the geodesic (see Appendix). Writing $ u^a_C = \dot t {\delta^a}_t + \dot r{\delta^a}_r$
\begin{equation}
\label{conserv}
(\pmb{\xi} {\cdot \bf u_C})= f\dot{t} = \epsilon = const.
 \end{equation}
 where $\epsilon$ determines from where C starts to fall.
Making use of (\ref{norm}) and (\ref{conserv}) the 4-velocity of $u_C$ can be written as:
\begin{equation}
\label{matrix}
{\bf u}_C = 
\begin{pmatrix}
\dot{t}\\
\dot{r} \\
\end{pmatrix}
 =
 \begin{pmatrix}
{\epsilon f^{-1}}\\
{[(\epsilon^2 f^{-1} - 1)/h]^{1/2} }\\
\end{pmatrix}
 \end{equation}
Now suppose that clock C starts to fall from the place $r = r_C > r_A > r_B $, we calculate the scalar product between the 4-velocities ${\bf u}_C $ with ${{\bf u}_A}$ when C reaches A 
\begin{equation}
\label{velA}
{({{\bf u}_A} \cdot {\bf u}_C )}|_{r_A}  = - \epsilon f^{- 1/2} (r_A)
  \end{equation}
and similarly for C with B, to give
\begin{equation}
\label{ratio2}
\frac{({\bf u}_B \cdot {\bf u}_C )| _{r_B}}{( {\bf u}_A \cdot{\bf u}_C )| _{r_A}} =  \Bigl (\frac{f(r_A)}{f(r_B)}\Bigr)^{1/2}= \left.\frac{d\tau_A}{d\tau_B}\right|_G
 \end{equation}
where the subscript G refers to gravity while M will be used for the corresponding expression  in Minkowski space.
Thus the ratio of the clock rates between A and B can be inferred by comparing the relative velocities with respect to C  at encounter. 
Although the velocities depend on the height from which C starts to fall, the ratio of the clock rate is independent. The reason is that spacetime is static and there is a simple geometrical argument which is given in the Appendix.
Now we consider clocks in Minkowski space.
The freely falling clock C becomes  a clock in uniform motion and for simplicity one may choose a system where C is at rest, while A and B undergo accelerations (Fig. 2). Then applying (\ref{rates1}) one has 
\begin{equation}
\label{ratio3}
\left.\frac{d\tau_A}{d\tau_B}\right|_M = \frac{ (1 - v_B^2)^{1/2}}{ (1 - v_A^2)^{1/2}}
\end{equation}
To reproduce the gravitational result for a general metric (\ref{metric1}) we just need to arrange the velocities of A and B in such a way that at encounter with clock C the relation (\ref{ratio2}) is satisfied.
\subsection{Constant Acceleration}
 So far we have only trivially adjusted the velocities of A  and B at encounter with C, but not specified the acceleration of how to achieve this.
In the gravitational setting the clocks A and B sit at fixed positions  in the static field and experience  a constant acceleration.
For a particle in Minkowski space with acceleration parallel to its velocity the relation between the proper acceleration $\alpha$ and the 3-velocity $v$  is given by
\begin{equation}
\label{alpha}
\alpha = \gamma^3 (v)\frac{dv}{dt}.
\end{equation}
If $\alpha$ is constant one may integrate (\ref{alpha}) in 1+1 dimension leading to hyperbolic motion (we use  coordinates $t$ and $x$ in Minkowski space)
\begin{equation}
\label {accel1}
(x(t) - \lambda)^2 - t^2 = \frac{1}{\alpha^2},
\end{equation}
where $\lambda$ is an integration constant and $t$ has been chosen so that the velocity is zero for $t = 0$. This implie that $ \lambda = x(0) - 1/\alpha $.
From (\ref{accel1}) we see that  the velocity can be written as
\begin{equation}
\label {vel1}
v =  \frac {dx}{dt} = \frac{t}{x -\lambda} = (x - \lambda)^{-1} \left((x - \lambda)^2 - \frac{1}{\alpha^2}\right)^{1/2},
\end{equation}
and thus,
\begin{equation}
\label {vel2}
(1 - v^2) = \alpha^{-2}(x - \lambda)^{-2}.
\end{equation}
 We now assume clocks A and  B to move with constant proper accelerations along the  trajectories
$x_A(t)$ and $ x_B(t)$
\begin{equation}
\label {accel2}
(x_A(t)  - \lambda_A)^2 - t^2 = \frac{1}{{\alpha_A}^2},
\end{equation} 
and correspondingly for B. 
We assume that both clocks start to accelerate at the same time with respect to the inertial clock C (this can always be achieved if the spacetime distance of their starting points is spacelike).
In order that the trajectories do not cross we choose $\lambda_A > \lambda_B$  and $x_A(0) > x_B(0) $. 
The clock rates of A and B are compared to the rate of the inertial clock C at $x_C  > x_A(0)$ when they meet i.e. $x_A(t) = x_C$ and $ x_B(t) = x_C $ .
From (\ref{vel2}) we see that the ratio of the proper times of A and B at the location of C is given by
\begin{equation}
\label {ratio2a}
\left.\frac{d\tau_A}{d\tau_B}\right|_{x_C} =  \Bigl (\frac{1  - v^2_A (x_C)}{1  - v^2_B(x_C)}\Bigl )^{1/2} = 
\frac{\alpha_B (x_C - \lambda_B)}{\alpha_A (x_C - \lambda_A)}.
\end{equation}
Thus the ratio of the clock rates depends in general  on the location of clock C . 
When both accelerations are equal i.e. setting  $\alpha = \alpha_A  =  \alpha_B$, the ratio is
\begin{equation}
\label {ratio3}
\left.\frac{d\tau_A}{d\tau_B}\right|_{x_C}  = 
\frac{x_C - \lambda_B}{ x_C - \lambda_A}.
\end{equation}
If clock A starts to accelerate from the location of clock C, i.e. $ x_A(t = 0) = x_C $ then, inserting for ${\lambda}_{A,B}$ in terms of $x(0)_{A,B}$,
(\ref{ratio3}) reduces to
\begin{equation}
\label {ratio4}
 \left.\frac{d\tau_A}{d\tau_B}\right|_{x_C}  = 
\alpha (x_C - \lambda_B) = 1 +{( g/c^2)} \Delta l,
\end{equation}
where
\begin{equation}
\label {approx}
\Delta l = x_A(0) - x_B(0)\qquad  \text{and} \qquad g ={\alpha} c^2
\end{equation}
where we have explicitly displayed the velocity of light $c$.
In this special case the result is the same as the one for clocks located at a height  difference $ \Delta l$  in a weak gravitational field of acceleration g. 
For  $\alpha_A$ different from $\alpha_B$
  the ratio of the clock rates will only be independent of the location of the reference clock C, if $\lambda_A = \lambda_B$,
\begin{equation}
\label {ratio5}
\left. \frac{d\tau_A}{d\tau_B}\right|_M = \frac{\alpha_B}{\alpha_A},
\end{equation}
similarly to the gravitational case where the ratio is independent of the height from which clock C starts to fall. Given two clocks resting at different height in a static gravitational field (1+1 dimensions), there always exist two uniformly accelerated clocks in Minkowski space so that the ratio of their proper times read off when passing the same, but arbitrary point in \it space\rm, equals the proper time ratio of corresponding stationary  clocks in the gravitational field.  
\subsection{Rigid Motion}
In our example of the two static clocks 
 it is tacitly assumed that the distance between the clocks remains constant. But as is well known, spatial distances in relativity are observer- dependent. So what does it mean that a body is in rigid motion?  
For special relativity the answer was already given by Born  (\cite{Born}) in 1909 and shortly after generalized by Herglotz  \cite{Herglotz} and Noether \cite{Noether}. The Born definition, which also carries over to general  relativity, says that rigid motion implies that the distance of neighboring particles measured in the momentary rest frame remains constant (see \cite{Trautman} and  also \cite {Giulini2} in the context of Minkowski space and references therein). 
 We give the mathematical formulation of this in the Appendix. 
It is clear that clocks resting in a static gravitational field are in rigid motion. In Minkowski space any motion that is a symmetry is a rigid motion, among them constant rectilinear acceleration \cite{Rindler}. In the considered case only if $\lambda_A = \lambda_B$  the two clocks are in rigid motion with respect to each other and keep a constant proper distance. But note that rigid motion implies that A and B have to undergo constant but different accelerations.
\subsection{Equivalence Principle}
Let us see what this means for the EP as Einstein formulated it. Einstein had in mind a uniform field i.e.  a static field where the acceleration is everywhere the same. 
In GR the 4-acceleration of a body is measured by its deviation from the geodesic motion.
For a static clock in the gravitational field  (\ref{metric1}) the absolute value of its 4-acceleration  $\bf a$ is  given by the expression (we give an explicit derivation in the Appendix)
\begin{equation}
\label {accel4}
|{\bf a}| = | g_{ab} a^a a^b |^{1/2} =  \frac{1}{2}\frac{f'}{f\sqrt{h}},
\end{equation}
where prime refers to the derivative with respect to $r$. In the 1+1 dimensional case it is useful to redefine the $r$-coordinate so that $h = 1$ to simplify  calculations. Note that this gauge choice does not alter the geometric conclusions.  
If the gravitational acceleration is independent of the height at which the clocks are situated, then 
\begin{equation}
\label {f4}
f'/2f = g = const.,
\end{equation}
( $h(r) = 1$ ), integration gives 
\begin{equation}
\label {f5}
 f(r) =C e^{2gr}\qquad    C = const.,
\end{equation}
and the ratio of the static clocks is
\begin{equation}
\label{ratio8}
\left. \frac{d\tau_A}{d\tau_B}\right|_G = \Bigl(\frac{e^{2gr_A}}{e^{2gr_B}}\Bigr)^{1/2} = e^{g(r_A - r_B)}.
\end{equation}
The situation in Minkowski space for equally accelerated clocks, $\alpha_A= \alpha_B  = g$ has already been considered, leading to (\ref{ratio3}) and in the special case to (\ref{ratio4}).
One sees that (\ref{ratio8}) agrees only to first order in $g \Delta l $, where $\Delta l$ is the proper distance between the clocks in both cases.
This is somewhat surprising: The ratio of the rates of two clocks resting at different heights in a static gravitational field of constant acceleration, cannot be simulated by clocks with constant and equal acceleration in Minkowski space. This contradicts Einstein's formulation that a \it uniform \rm  gravitational field is equivalent in regard to \it all \rm physical processes to a system in uniform acceleration. But at the time Einstein had in mind the uniform field in Newtonian gravity.
One may question whether the generalization of Newton's  uniform gravitational field  (\ref{f5}) is not too special. If one considers a general $f(r)$, then in order to reproduce the gravitational result with accelerated clocks in Minkowski  space, we have to require
 \begin{equation}
 \label {ratio6}
\left.\frac{d\tau_A}{d\tau_B}\right|_{M} = \frac{\alpha_B}{\alpha_A}  = \Bigl (\frac{f(r_A)}{f(r_B)}\Bigr)^{1/2}=\left.\frac{d\tau_A}{d\tau_B}\right|_{G}
  \end{equation}
 i.e. the accelerations $\alpha_A $ and $\alpha_B$ are related to the gravitational\it{ potential}  \rm and not to its gradient, and will not be equal to the gravitational acceleration as given by (\ref{accel4}). 
If we use these accelerations in the expression for the clock rates in Minkowski space  
 then 
\begin{equation}
\label {ratio7}
\frac{\alpha_B }{\alpha_A} =  \frac{f'(r_B) f(r_A)} {f'(r_A) f(r_B)}.
\end{equation}
Enforcing  this to be equal to (\ref{ratio6}) leads to the condition
\begin{equation}
\label {f1}
 f' f^{-1/2}|{r_A} =  f' f^{-1/2}|{r_B}
\end{equation}
Since this should be valid for arbitrarily chosen $r_A$ and $r_B$ we have
\begin{equation}
\label {f2}
 f'(r) f(r)^{-1/2}= b = const.
\end{equation}
which integrates to
\begin{equation}
\label {f3}
 f(r) = \bigl (\frac {br - C}{2}\bigr)^2
\end{equation}
where $C$ is an integration constant.
Thus, if we require that the accelerations of A and B are equivalent to the acceleration that the corresponding  clocks experience  in the static field, then the gravitational field is restricted  to (\ref{f3}). Calculating the curvature tensor for this metric shows that it is identically zero which implies that spacetime is flat but expressed in accelerated coordinates (Rindler \cite{Rindler}). 
This is not unexpected since tidal forces prevent accelerated clocks in Minkowski space to mimic the gravitational effects exactly. Moreover, as shown, even the effect of a gravitational field of everywhere equal magnitude cannot be simulated by equally accelerated clocks. Also, if one tries to generalize the concept of a uniform field to four-dimensions, the only vacuum solution to Einstein's equations is flat spacetime \cite{Aichelburg}. Thus, no analog to Newton's uniform field in GR exists. 
\section {FREQUENCY SHIFT}
 Already in 1907 Einstein \cite{Einstein1907} wrote  `` ...there are clocks whose rates can be controlled  with great precision, these are producers of spectral lines''  and concludes that the wavelength of light coming from the sun's surface should be ``larger by about one part in two millions''.  (For recent experimental verification of the gravitational redshift see \cite{Vessot}, and references therein).  In this respect frequency is directly related to the inverse of proper time and the results of the previous section can immediately be applied. Nevertheless let us see how the EP can be applied directly to light rays.  First, we give the qualitative argument:
Consider now two detectors A and B  (instead of clocks) vertically mounted in a static gravitational field.  A light ray is sent vertically past them where the frequency is measured. Invoking the EP in the gravity-free case,  the detectors are accelerated. Therefore the detectors will have different velocities as the ray passes and due to the Doppler shift register different frequencies.
In the gravity setting given by the metric (\ref {metric1}) the detectors rest in the static field  and the light ray follows a  geodesic. Therefore, the geometric arguments given in the Appendix apply and lead to
\begin{equation}
\label{freq1}
\left.\frac{\omega_B}{\omega_A}\right|_G = \frac{({\bf u} \cdot {\bf l})_B}{({\bf u} \cdot{\bf l})_A} = \Bigl(\frac{f(r_A)}{f(r_B)}\Bigr)^{1/2},
\end{equation}  
where $\omega_A$ and $\omega_B$ are the measured frequencies, ${\bf l}$  is the tangent null vector along the ray and the scalar product is to be evaluated at  $r_A$ and $ r_B$ respectively. As expected, this is the inverse ratio as for clocks rates.
How is this mapped into Minkowski spacetime?  
In Minkowski space the relativistic Doppler shift is given by
\begin{equation}
\label{freq2}
\bar \omega\ = \omega_0 \Bigl(\frac{1 + v}{1 -v}\Bigr)^{1/2},
\end{equation}
where $\bar\omega$ is the frequency measured by an observer moving with velocity $v$ towards the emitter of proper frequency $\omega_0$.  We need to know the velocities of the detectors when the ray passes. As for the clocks we consider the two detectors to be rigidly accelerated, while the light ray travels along a straight worldline. The trajectory of constant acceleration is again given by (\ref{accel1}) and  from  (\ref{vel1}) we find for $\lambda = 0$ 
\begin{equation}
\label {vel3}
\frac{ 1 + v}{1 - v} = \frac{x + t}{x - t},
\end{equation}
Consider a light ray propagating toward the accelerated detector whose trajectory is given by $x = x_0 - t $, where $x_0 > 1/\alpha$ .  When the ray reaches the detector the corresponding coordinates $ (x, t)$ agree and  from (\ref{accel1}) we get
\begin{equation}
\label {accel5}
(x - t)^{-1} = (x + t)\alpha^2 = x_0\alpha^2,
\end{equation}
and therefore
\begin{equation}
\label {vel4}
\frac{ 1 + v}{1 - v} =  (x_0\alpha)^2.
\end{equation}
If we apply this to the two detectors in the case  $\lambda_A  = \lambda_B $, 
(one may set both $\lambda$'s to zero by a simple translation in $x$-direction),
then we get  
\begin{equation}
\label {freq3}
\left.\frac{\omega_B}{\omega_A}\right|_M = \frac{x_0 \alpha_B}{x_0 \alpha_A},
\end{equation}
which is independent of the trajectory of  the ray and is the inverse of the corresponding clock rates.
 This shows that in order to reproduce the gravitational frequency shift the detectors have to be in relative rigid motion however, as discussed in the previous section, the required acceleration in Minkowski space will not agree with the acceleration of the static detectors. 
\section {LOCAL BENDING OF LIGHT RAYS}
The total deflection of a light ray passing close to a massive star is of course a non-local effect and the EP can hardly be valid. But instead of considering the total deflection one can look at the ``local" bending. The idea is to calculate the curvature of the \it spatial \rm orbit of the ray at a given point and compare it with bending in flat spacetime as seen by an accelerated observer. By orbit in this context we mean the space and not the spacetime trajectory (worldline).  Geometrically this means that the worline of the ray is projected onto the planes of constant Killing time (see Fig.  3).
\begin{figure}
 \centering
 \includegraphics[scale=0.40]{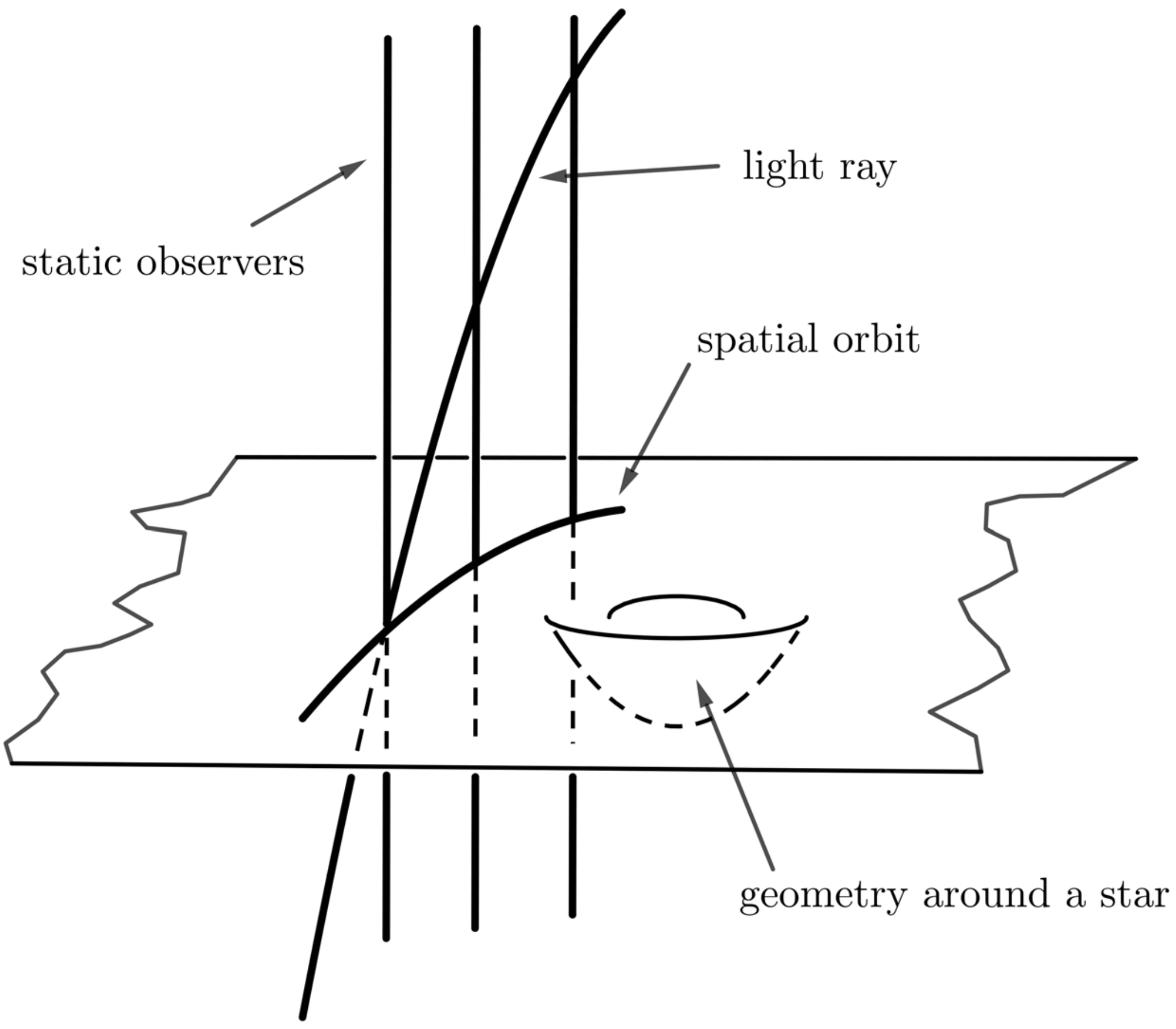}
 \caption{GEOMERTY OF LIGHT BENDING BY GRAVITY   the figure shows schematically a light ray in curved spacetime passing near a star.  The projection of the ray onto a surface orthogonal to the static observers traces out a spatial orbit.   
  }\label{}
 \end{figure}
Let us repeat the qualitative argument that can be found in many textbooks on GR:
Without gravity, consider an accelerated cabin with a light ray penetrating horizontally from one side (Fig. 4). During the time the light ray takes to reach the opposite wall, the cabin has moved, causing the ray not to hit exactly the opposite point. If the cabin is accelerated, then the ray will deviate from a straight line as viewed from the cabin. Applying the EP, the same light bending should occur in a cabin resting in a (static) gravitational field. (For a spacetime picture see (Fig. 5)).

\begin{figure}
 \centering
 \includegraphics[scale=0.40]{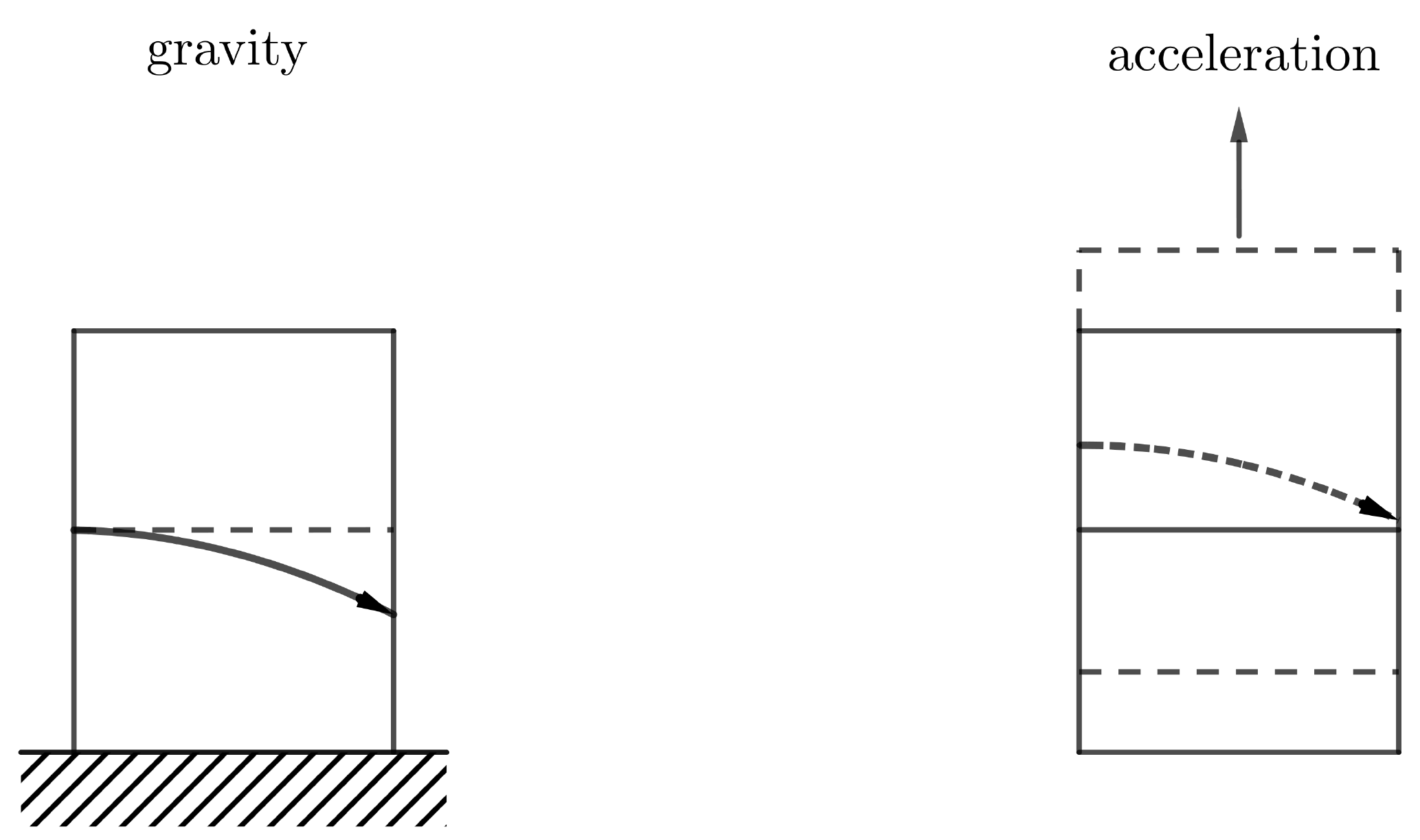}
 \caption{LIGHT BENDING BY GRAVITY  and as seen by an accelerated observer in Minkowski space (dashed curve).
  }\label{}
 \end{figure}

\begin{figure}
 \centering
 \includegraphics[scale=0.40]{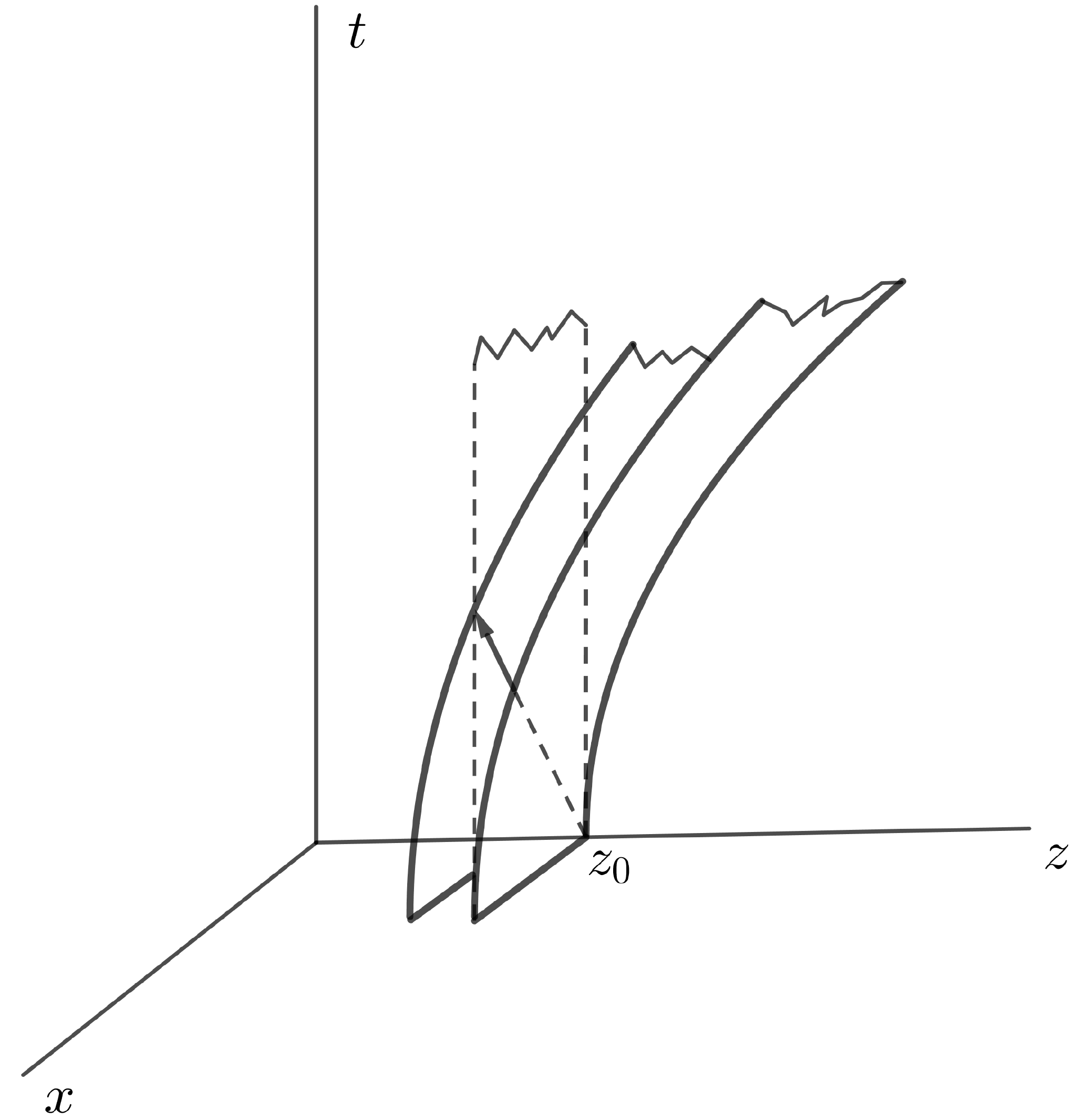}
  \caption{LIGHT BENDING FOR ACCELERATED OBSERVERS,
the figure shows two cross-sections of an accelerated cabin in Minkowski space.  A light ray in $x$-direction entering at $z_0$ stays in the plane $z = z_0$ , while ending up in a different cross section from the one in which it started. 
}\label{}
\end{figure}
We start by considering this effect in Minkowski space. For this we need two space directions, thus we consider an accelerated cabin in spacetime of 1 + 2 dimensions with coordinates  $(t, x ,z)$:
\begin{equation}
\label{metric2}
ds^2 = -dt^2 + dx^2 + dz^2.
\end{equation} 
Now the question arises how the cabin should be accelerated? As before, we assume constant acceleration, but since the cabin is an extended object, rigid motion is the natural requirement. (But see comment at the end of Sec. A.) 
\subsection{Accelerated Cabin}
The cabin should undergo rigid acceleration in the $z$-direction. Different cross sections of the cabin trace out a one-parameter congruence of 2-dimensional surfaces in spacetime (in 1+2 dimensions),
parametrized by $\alpha$.
\begin{equation}
\label {accel6}
z(t;\alpha)^2 - t^2 = 1/\alpha^2,\quad x(t;\alpha) = const.
\end{equation}
Points of equal $z$-coordinate experience the same acceleration, while the acceleration varies from point to point in the $z$-direction. 
For the sake of simplicity we assume that the light ray enters at the instant when the cabin is momentarily at rest, traveling along the positive $x$-direction i.e. $t = 0$   with $x(0) = 0, z(0) = z_0$
\begin{equation}
\label {ray1}
x = t , \qquad z  = z_0.
\end{equation} 
We introduce new coordinates  $(T,X,Z)$ so that $Z$ is constant along the accelerated trajectories (one could also transform to Rindler co-moving coordinates \cite{Rindler}, which however would lead to the same result),
\begin{equation}
\label {coord1}
T =  t, \quad X = x, \quad Z^2 = z^2 -t^2
\end{equation}
Then the light ray in the new coordinates is given by
\begin{equation}
\label {coord2}
Z^2 = z_0^2 -T^2, \qquad X = T.
\end{equation}
Thus
\begin{equation}
\label {coord3}
Z =( z_0^2 -X^2)^{1/2},
\end{equation}
and the local bending turns out to be 
\begin{equation}
\label {orbit1}
\frac {d^2 Z}{dX^2} = - z_0^2 (z_0^2 - X^2)^{-3/2}.
\end{equation}
At $X = 0$, where the ray enters the cabin orthogonal to the acceleration we have
 \begin{equation}
 \label {orbitA}
\left.\frac {d^2 Z}{dX^2}\right |_{X=0} = -1/z_0 = - \alpha_0,
\end{equation}
where $\alpha_0$  is the proper acceleration of the cabin at the point $z_0$.
Transforming the metric (\ref{metric2}) to the co-moving coordinates $T, X, Z$ leads to 
\begin{equation}
\label {metric3}
ds^2 = - dT^2 +dX^2 + \frac {(TdT + ZdZ)^2}
{T^2 + Z^2}
\end{equation}
Since for $T=0$ the spatial metric is flat, we may  calculate the first curvature $\kappa$ of the curve $Z(X)$ making use of the standard formula (for consistency in later application we take as definition for $\kappa$ the absolute value of the second derivative along the curve).
\begin{equation}
\label {curv1}
\kappa=\left|\frac{d^2Z}{dX^2}\right|\Bigl (1 + \Bigl (\frac{dZ}{dX} \Bigr)^2 \Bigr )^{-3/2}
\end{equation}
evaluated  at $X = 0$  gives (\ref{orbitA}).
This shows that an accelerated observer in Minkowski space will measure the curvature of a light ray  to be equal to its acceleration (in units where $c = 1$). 
Calculations where all points of the cabin undergo accelerations of the same magnitude follow along the same line leading to the same curvature  (left as an exercise to the reader). This not surprising since the difference between rigid and equal accelerations is expected to shows up only in non-local effects.

\subsection{Bending by Gravity}
Now we look at the local bending of light in a gravitational field. Consider the static metric in 2+1 dimensions of the form
\begin{equation}
\label{metric3}
ds^2 = -f(z)dt^2 + dx^2 + dz^2
\end{equation}
For the trajectory of a light ray we have $ds^2=0$,
\begin{equation}
\label {norm1}
-f(z)\dot t^2 + \dot x^2 +\dot z^2 = 0,
\end{equation}
where the dot now denotes the derivative with respect to an affine parameter along the ray. Time- and $x$-translational invariance leads to  
\begin{equation}
\label {const1}
f(z)\dot t = C_1,  \qquad\dot x = C_2,
\end{equation}
where $C_1 $ and $ C_2 $ are integration constants. Inserting this into (\ref{norm1})  gives
\begin{equation}
\label {const2}
f(z)(\dot z^2 +C_2^2) = C_1^2.
\end{equation}
Since $x$ depends linearly on the affine parameter, we may write 
\begin{equation}
\label {const3}
\dot z = \frac {dz}{dx}\dot x = \frac {dz}{dx} C_2,
\end{equation}
to get $z(x)$
\begin{equation}
\label {const4}
f(z)\Bigl (\left(\frac{dz}{dx}\right)^2 + 1\Bigr ) = \Bigl (\frac {C_1}{C_2}\Bigr )^2.
\end{equation}
If we choose as initial conditions $ (dz/dx) = 0$  at $ z = z_0 $  i.e. the ray enters the cabin at $ z_0$ in the positive $x$-direction, orthogonal to the acceleration of gravity, then
\begin{equation}
\label {const5}
\Bigl (\frac {C_1}{C_2}\Bigr )^2 = f(z_0),
\end{equation}
and
\begin{equation}
\label {bend1}
\frac {dz}{dx} = \Bigl (\frac{f(z_0)}{f(z)} - 1\Bigr )^{1/2}.
\end{equation}
The second derivative of $z$ with respect  to $x$ gives the local bending of the ray
\begin{equation}
\label {bend2}
\frac {d^2z}{dx^2} = - \frac{f(z_0)}{2f(z)^2} \frac {df}{dz}. 
\end{equation}
Since the spatial slices of the metric (46) are flat, the curvature $\kappa $ of the orbit is simply given by the absolute value of the above expression evaluated at $z= z_0$,
\begin{equation}
\label {curv2}
\kappa|_{z_0} =
 \left.\frac{1}{2f(z)}\frac {df}{dz}\right|_{z_0}. 
\end{equation}
This in turn is equal to the 4-acceleration of a static observer (\ref{accel1}), at $ z = z_0$. 
How does this compare with the local bending of a light ray seen by a uniformly accelerated observer in Minkowski space?  As shown, at the point where the direction of the ray is orthogonal to the acceleration of the cabin, the resulting curvature of the ray is equal to the acceleration (divided by $c^2$) and is therefore equal to the bending by the gravitational field.  
\subsection{Local bending in the Schwarzschild Spacetime}
Now we turn to a more physical spacetime: the gravitational field produced by a spherically symmetric mass, the famous Schwarzschild metric.
\begin{equation}
\label{SS1}
ds^2 = - Adt^2 + A^{-1} dr^2 +r^2 d{\phi}^2, \quad A =1 - 2GM/r
\end{equation}
where  without loss of generality, we have set $\theta  = \pi/2$,  and consider only rays in that plane. 
The acceleration of a static observer can again be obtained from (\ref{accel4}), yielding
\begin{equation}
\label {accel7}
|{\bf a}| = \frac{1}{2}A' A^{-1/2} = \frac{GM}{r^2} A^{-1/2}. 
\end{equation}
The analysis of the propagation of a light ray in the metric (\ref{SS1}) may be found in any textbook on GR (see e.g. \cite {Rindler2001}). The spatial orbit $r(\phi)$ satisfies
\begin{equation}
\label {orbit2}
 \frac{d^2r}{d\phi ^2} = \frac{2}{r}\left(\frac {dr}{d\phi}\right)^2 + r - 3GM.
\end{equation}
If the EP is valid, there should be a relation between the acceleration experienced by a static observer and the bending of the ray. As before, we look at the bending where $({dr/d\phi})|_{r_0} = 0$  and the ray trajectory is orthogonal to the gravitational acceleration i.e. at perihelion. Then
\begin{equation}
\label {orbit3}
\left.\frac{d^2r}{d\phi ^2}\right |_{r_0} =  r_0 - 3GM.
\end{equation}
Although the coordinates  $r$ and $\phi$ have a geometrical meaning, one would like to have a coordinate invariant expression for the curvature. Since the 2-dimensional subspace $t = const.$ of the metric is not flat, we cannot simply apply the formula of Euclidean space. Following Eisenhart \cite{Eisenhart}  the curvature of a curve  in Riemannian space is given by its principal normal 
\begin{equation}
\label {bend3}
k^i = t^j D_j t^i,
\end{equation}
where $\bf t$ is tangent to the curve and the covariant derivative $D$ in our case refers to the induced metric $h_{ij}$ on $t= const.$ slices (see eq. \ref{rigid}).  
 The curvature $\kappa$ of $r(\phi)$ is given by the absolute value of $k^i$ (indices refer to $(r, \phi)$).
\begin{equation}
\label{bend4} 
 \kappa =|{\bf k}| = |g_{ij}k^ik^j|^{1/2}.
\end{equation}
 Given the spatial metric
\begin{equation}
\label {SS2}
dl^2 =  A^{-1} dr^2 +r^2 d{\phi}^2,
\end{equation}
we consider first  an arbitrary orbit $r(l)$, $\phi (l) $, parametrized by the arclength of the curve $ l$, therefore
\begin{equation}
\label {norm2}
1 =   A^{-1} r'^2 +r^2 {\phi'}^2,
\end{equation}
with a prime denoting the derivative with respect to $l$.
Calculating the curvature of the orbit according to (\ref{bend3}) at the point where $r' = 0$ results in ( for details see the Appendix)
\begin{equation}
\label {bend5}
k^r|_0 = r''_0 - A_0r_0{\phi'_0}^2\quad{k^{\phi}}|_0 = {{ \phi}''}_0
\end{equation}
From (\ref{norm2}) we infer that at $ r_0$ that  $(\phi')_0 = 1/r_0$ and after differentiation that $\phi''_0 = 0$. Taking the absolute value of $\bf k$ gives
\begin{equation}
\label {bend6}
\kappa|_0 = A^{-1/2}(|r'' - A/r|)|_0.
\end{equation}
Now we consider the spatial orbit of a light ray i.e. its projection onto a  $t = const.$ surface. To make use of (\ref{bend6}), we need to translate the derivatives with respect to $\phi$ into
 those with respect to $l$, 
\begin{equation}
\label {orbit4}
\frac{dr}{d\phi} = \frac{dr}{dl}\cdot\frac{dl}{d\phi} = r'/\phi',
\end{equation}
which yields 
\begin{equation}
\label {orbit5}
\frac{d^2r}{d\phi ^2} =  \frac{1}{(\phi')^2}\left(r'' - \frac{r'}{\phi'}\phi''\right).
\end{equation}
Evaluating at $ r' = 0$,
\begin{equation}
\label {orbit6}
\frac{d^2r}{d\phi ^2}|_0 =    \frac{r''}{(\phi')^2}|_0 = r''r^2|_0.   
\end{equation}
Substituting for $r''$ in (\ref {bend6}) from (\ref{orbit6}) and making use of (\ref{orbit3}) 
we find
\begin{equation}
\label {orbit7}
\kappa|_0 = \frac{GM}{r^2}A^{-1/2}|_0,
\end{equation}
and the value agrees with (\ref {accel7}).
We conclude that the curvature of the \it spatial \rm orbit of a light ray in the considered static gravitational field (here specifically in the Schwarzschild spacetime)  is equal to the acceleration of a static observer at that point. As shown for the accelerated cabin, a light ray in Minkowski space will show the same curvature as seen by an observer subjected to the equivalent acceleration as the static observer.
In this sense the EP is confirmed without assuming the gravitational field to be weak. Note that to first order in $GM/r$ eq.(\ref{orbit7}) is just the Newtonian value.
 It has been argued that the EP contributes only one half to the total deflection of the ray, the other half coming from the space curvature. The authors \cite{Ferraro, Moreau} tried to disentangle these two effects. Our derivation for the Schwarzschild metric however shows that even for local light bending the space curvature plays a role.  An effect which however does not contribute in the weak field approximation.  It may be verified that modifying the Schwarzschild metric to have flat space slices, would lead to an additional factor  $A^{1/2}$ in (\ref{orbit7}).
 But then the  acceleration of a static observer changes accordingly. Thus the above statement about the equivalence of the local light bending in a static field and the corresponding accelerated observer remains valid.
\section{TIME DELAY AND THE EP}
Consider the metric for a general spherically symmetric and static gravitational field.
\begin{equation}
\label {metric4}
ds^2= -f(r)dt^2 + h(r)dr^2 + r^2d\Omega^2:
\end{equation}
To simplify we look only at a radial light ray  starting at $r = r_2$ to be reflected at $r_1 < r_2$ and back to $r_2$  and calculate the proper travel time as measured by an observer at $r_2$ (see Fig.6).
\begin{figure}
 \centering
 \includegraphics[scale=0.50]{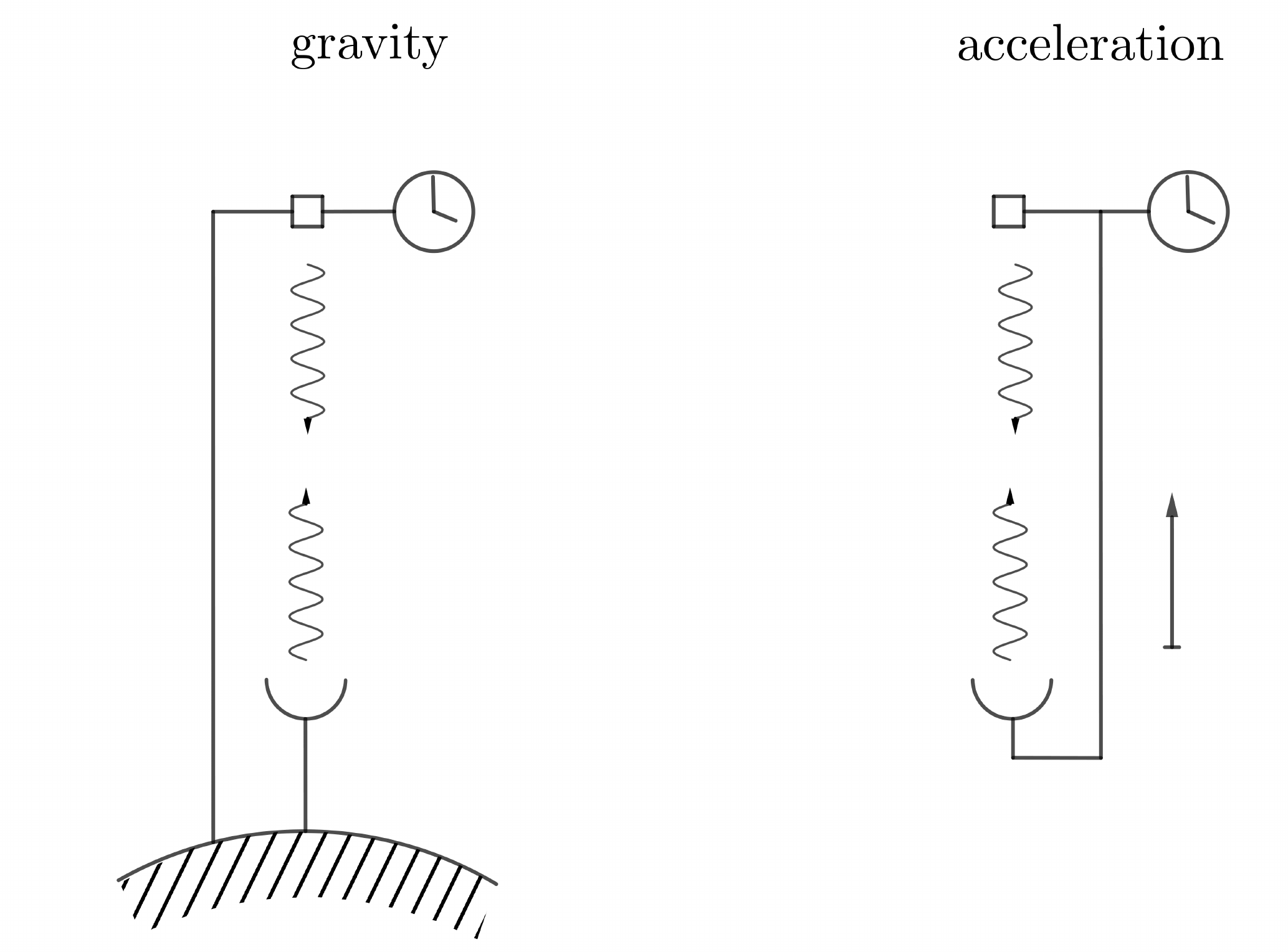}
 \caption{BOUNCING RAY, a ray is sent radially towards a reflector in a gravitational field,  bounces back and local travel  time is recorded (l.h.s.) A similar arrangement is accelerated in Minkowski space (r.h.s.).
  }\label{}
\end{figure}
From $ds^2= 0$
we have
\begin{equation}
\label {vel5}
\frac{dr}{dt} = \pm\Bigl (\frac{h(r)}{f(r)}\Bigr )^{-{1/2}}, 
\end{equation}
the elapsed coordinate time for the ray from $r_1$ to $r_2$ is
\begin{equation}
\label {time1}
\Delta t = \int _{r_1}^{r_2} \Bigl (\frac{h(r)}{f(r)}\Bigr )^{1/2} dr,
\end{equation}
while the proper time at $r_2$ is given by
\begin{equation}
\label {time2}
\Delta \tau =  f(r_2)^{1/2}\Delta t =   f(r_2)^{1/2} \int _{r_1}^{r_2} \Bigl (\frac{h(r)}{f(r)}\Bigr )^{1/2} dr.
\end{equation}
If, as in Sec. II, we assume that  $f' > 0$, then
\begin{equation}
\label {time3}
\Delta \tau >  \int _{r_1}^{r_2 }h(r)^{1/2} dr = \Delta l,
\end{equation}
but the integral is just the proper (physical) distance $\Delta l$ between $r_1$ and $r_2$, therefore
$\Delta \tau  > \Delta l$.
A ray in the opposite direction will take the same time, and the total time for the bouncing ray is twice this value. Without gravity $f(r) = h(r) = 1$ we simply have $\Delta \tau = \Delta l$. 
Thus, for the static observer the bouncing ray will take longer proper time for the same distance in a gravitational field than in flat spacetime.
What does this have to do with the EP?
If we set $ f(r) = r^2$ and $ h(r) = 1$, the metric resembles  flat space in Rindler \cite{Rindler} coordinates.  This means that observer and mirror with $r = const. $  are inuniform acceleration  in Minkowski space. They are in rigid motion and thus keep constant distant to each other. Explicitly we have
\begin{equation}
\label {time4}
\Delta \tau = r_2\ln (r_2/r_1)> r_2 - r_1 = \Delta l
\end{equation}
$(c=1)$, which shows that there is a time delay of signals for rigid accelerated observers in Minkowski space.  However, the magnitude of the time delay will not coincide with the one in gravity. As argued, the acceleration cannot mimic the changes of the gravitational field along the path of the ray.  
\section{CONCLUSION}
We have analyzed if the effects of gravity on clock rates and the propagation of light rays  can be reproduced in Minkowski spacetime as seen by a suitably accelerated observers. We assumed that gravity is described by a spacetime metric and restricted ourselves to static spacetimes. Moreover, freely falling clocks and light rays were assumed to follow geodesics of the given metric. No assumptions were made about the gravitational strength i.e. we did not require the field to be weak nor did we consider the Newtonian limit. The main results can be summarized as follows:
i) Given two clocks resting at different heights in a static gravitational field (1+1 dimensions), there always exist two uniformly accelerated clocks in Minkowski space so that the ratio of their proper times read off when passing the same but an arbitrary point in space, equals the proper time ratio of corresponding stationary  clocks.  If in addition one imposes that the accelerations are equivalent to the acceleration that the static clocks experience, then this restricts  spacetime to be flat. ii) The ratio of the rates of two clocks resting at different heights in a static gravitational field of constant acceleration cannot be simulated by constant and equally accelerated clocks in Minkowski space.
iii)
Local light bending in static spherically symmetric spacetimes can be simulated by accelerated observers in Minkowski space despite the fact that the bending depends (also) on the curvature of the spatial subspace. This holds also for strong gravitational fields.
iv) The effect of time delay of light rays also exists for rigidly accelerated observers in Minkowski space. But this covers only part of the delay in a gravitational field since the curvature of spacetime also contributes to the effect. 
So is the EP useful to understand the effects of gravity? The answer is ambivalent: on the one hand, it may serve as a tool for understanding which effects caused by gravity on physical systems are of inertial nature, i.e. which effects  can be simulated by acceleration. For example it tells why astronauts are weightless despite the existence of tidal forces. 
On the other hand, care must be taken when applying it to simple qualitative arguments which, as shown, do not find their mathematical correspondence.
Of course, this is all encoded in the theory of GR. Since a \it uniform \rm gravitational field does not exist in Einstein's theory, in any ``true'' gravitational field there are  tidal  forces acting. But this does not make the concept of the EP useless for understanding some of the implications of  GR.
\section{appendix}
\appendix*   
i) We give a simple argument leading to eqs. (\ref{conserv}) and  (\ref{ratio2}). Since the spacetime metric (\ref{metric1}) is static there exists a timelike Killing vector field  $\pmb{\xi}$
, which simply is $\pmb{\xi} = \partial/\partial t$,
satisfying
\begin{equation}
\label{kill1}
 \nabla_a\xi_b + \nabla_b\xi_a = 0,
  \end{equation} 
The static clocks A and B follow Killing trajectories
\begin{equation}
 \label{kill2}
{\bf u}_{A,B}= \left.\frac{ \pmb{\xi}}{| \pmb{\xi}|}\right|_{r_A,_B}, \qquad |\pmb{\xi}| =  f(r)^{ 1/2} ,
  \end{equation}
while C follows a geodesic i.e.
\begin{equation}
\label{kill3}
{ {u^a}_C\nabla_a{u^b}_C} = 0,
\end{equation} 
therefore
\begin{equation}
\label{kill4}
 {u^a}_C\nabla_a({\pmb \xi}\cdot{{\bf u}_C)} =   {u^a}_C\nabla_a(\xi_b){{u^b}_C} = 0. 
\end{equation}
Thus, the scalar product between the Killing vector (field) and the 4-velocity of a geodesic is constant along the geodesic, and we conclude
\begin{equation}
 | \pmb{\xi}|{({\bf u}_A \cdot{{\bf u}_C)}|_{r_A}}=| \pmb{\xi}|({{\bf u}_B \cdot{{\bf u}_C)}|_{r_B}},
\end{equation} 
and 
\begin{equation}
\label{kill5}
\frac{({\bf u}_B \cdot {\bf u}_C )| _{r_B}}{( {\bf u}_A \cdot {\bf u}_C )| _{r_A}} =
\frac{|\pmb{\xi}(r_A)|}{|\pmb{\xi}(r_B)|}.
\end{equation}
Since the norm of $\pmb{\xi}$  is  $f(r)^{1/2}$, this results in (\ref {ratio2}).

ii) In the following we give a brief derivation of eq.(\ref{accel4}). Because $ \pmb{\xi}$ is Killing
\begin{equation}
\label{X}
\xi^a\nabla_a(\xi^2) = 2 \xi^a(\nabla_a\xi_b)\xi^b = 0,
\end{equation}
and making use of (\ref{kill2}) allows us to write 
\begin{equation}
u^a\nabla_a u^b = -\frac{ 1}{\xi^2} \xi^a\nabla_a \xi^b = \frac{1}{\xi^2} \xi^a\nabla^b \xi_a =
 \frac{1}{2\xi^2}\nabla^b \xi^2 = \frac{1}{2f} g^{bc}\nabla_c f = \frac{ f'}{2f h}\delta^b_r,
\end{equation}
iii) Rigid motion in mathematical terms implies that (see \cite{Trautman})
\begin{equation}
\label{rigid}
L_u h_{ab} =0,  \qquad h_{ab} = g_{ab} + u_au_b,
\end{equation}  
where $ L_u$ is the Lie-derivative in the direction  $\bf u$ which is tangent to the world-lines of the rigid motion and $h_{ab}$ is the induced metric orthogonal to $\bf u$. If the motion is along a Killing trajectory, then (\ref{rigid}) is trivially satisfied.
 In  Minkowski space, the flow of an isometry (symmetry) is either i) rectilinear motion with constant velocity or ii) motion with uniform angular velocity or rectilinear uniform acceleration. One may wonder why the geometrical argument given above does not apply for the clocks in Minkowski space, i.e.\ why  the ratio  of the clock rates in Minkowski space is not independent of the site?  Since the trajectories of uniformly accelerated clocks follow integral curves of a ``boost" Killing field and clock C being inertial thus follows a geodesic. The answer is that although the accelerated clocks move on Killing trajectories,  A and B move along different Killing congruences for different  $\lambda$'s  which can be represented as a combination of the boost-Killing field $ x\partial_t + t\partial_x $ and a time translation $\partial_t$ i.e. $ \pmb{\xi}_\lambda = (x - \lambda)\partial_t + t\partial_x$.
 
iv) Finally we give hints to obtain eq.(\ref{bend5}). Writing eq.(\ref{bend3}) in the form
$k^a={ t^a}'' + {\Gamma^a}_{b c}{t^b}'{ t^c }'$,
with 
$t^a = r'\delta^a_r + {\phi}'{\delta}^a_{\phi}$
as the tangent vector to the curve $r(l), \phi(l)$ and making use of the condition  $r'_0=0$, which strongly simplifies the calculations, yields
\begin{equation}
{k^r}|_0 = r_0'' + {\Gamma^r}_{\phi \phi} {(\phi')}^2|_0, \qquad        
{k^\phi}|_0 = \phi''|_0
 \end{equation}
 with $ {\Gamma^r}_{\phi \phi}|_0 = - A_0 r_0$,
 as the only relevant non-zero $\Gamma$- coefficient of the metric (\ref{bend3}), gives eq.(\ref{bend5}).
\begin{acknowledgments}
The author thanks H. Balasin and C. Spreitzer for critical reading of the manuscript and for helpful comments. C. Spreitzer also provided the figures which is especially acknowledged.  
\end{acknowledgments}

\maketitle 

\end{document}